\documentclass[11pt,notitlepage]{isasrep}
\usepackage[dvips]{graphicx}
\usepackage[small,hang]{caption2}
\begin{document} 

\title{ Morphological Number Counts of Galaxies \\ 
  in the Hubble Deep Field South}
				

\markright{Hubble Deep Field South}


\author{
     {\sc Myung Gyoon Lee}\footnotemark[1]  $\;$ and $\;$
      {\sc Narae Hwang}\footnotemark[1]$\;$
} 


\date{(16 July 2000)} 

\maketitle	


\renewcommand{\thefootnote}{\fnsymbol{footnote}} 

\footnotetext[1]{Astronomy Program, SEES, Seoul National University,   KOREA
 (mglee@astrog.snu.ac.kr) }

\renewcommand{\thefootnote}{\arabic{footnote}}   
\setcounter{footnote}{0}			 
\renewcommand{\baselinestretch}{1}		 


\begin{abstract}
We present a study of photometric properties of the galaxies in the Hubble
Deep Field South (HDFS) based on the released WFPC2 images obtained with
the Hubble Space Telescope (HST). 
We have classified about 340 galaxies with $I<26$ mag in the HDFS as well
as about 400 galaxies in the Hubble Deep Field North (HDFN) using the
visual classification supplemented by inspection of the surface brightness
profiles of the galaxies.
Galaxy population statistics and morphological number counts for the HDFS are 
found to be similar to those for the HDFN. 
We have also determined photometrically the redshifts of the
galaxies with $I<26$ mag in the HDFS and the HDFN using the empirical training set
method.
Redshift distribution, color--redshift relation, and magnitude--redshift for each
type of galaxies are investigated.

\end{abstract}


\section{INTRODUCTION}

Hubble Deep Field Program has provided us with the deepest images presently possible
which are very useful as windows to understand the formation and evolution of
the distant universe (Ferguson, Dickinson \& Williams 2000). 
Following the Hubble Deep Field North (HDFN), the Hubble
Deep Field South was observed with WFPC2, NICMOS and STIS in October 1998
(Williams et al. 2000). We have analyzed the WFPC2 data among them, 
and have classified bright galaxies with $I<26$ mag. 
Here we present a preliminary study of
various aspects of the galaxies in the HDFS including morphology, number counts, 
redshift, and colors in comparison with the HDFN.

\section{PHOTOMETRY}

We have used a galaxy photometry software FOCAS (Faint Object Classification and Analysis System) (Valdes et al. 1995) to detect the objects 
and to obtain the photometry of the detected objects in the WFPC2 images. 
We have combined $B$, $V$ and $I$ images released to the community 
to create a master image which is used for final object detection with FOCAS.
Williams et al. (1996) also used FOCAS for the photometry of the galaxies
in the HDFN, while Williams et al. (2000)  used the Source Extractor (SExtractor)
(Bertin \& Arnouts 1996) for the photometry of the galaxies
in the HDFS. Details of the photometry are described in Hwang (2000).
Finally, we have obtained the photometry of about 2400 objects reaching
$V\sim 30$ mag in the HDFS, and of about 2700 objects in the HDFN.
A comparison of our photometry with the photometry given by \cite{wil00} 
shows that they agree well in general
and that the FOCAS magnitudes get somewhat brighter than the SExtractor magnitudes
at the faint level ($V>28$ mag). 

Fig. 1 displays an $I$ vs. $(V-I)$ color-magnitude diagram of the measured objects 
in the HDFS. Several features are seen in Fig. 1:
galaxies start to appear at $I\sim 21$ mag;
most galaxies are bluer than $(V-I) \sim 1.0$; and
galaxies get bluer progressively as they get fainter.
These features are, in general, similar to those for the HDFN. 

\begin{figure}[ht] 
   \begin{center}
   \includegraphics[angle=0,width=8cm,clip]{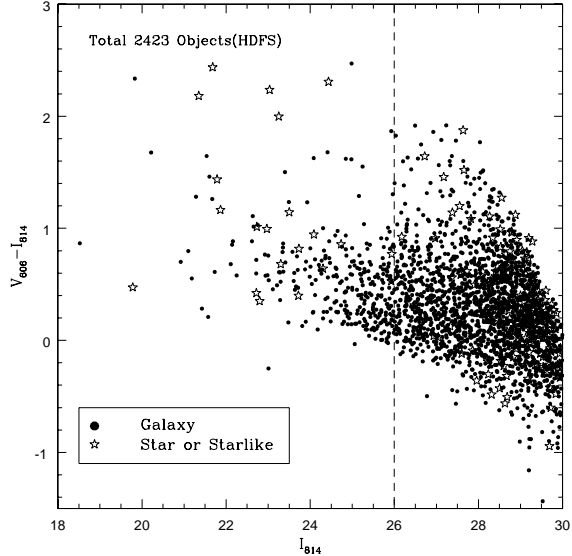}
   \caption{$I$ vs. $(V-I)$ color-magnitude diagram of the measured objects 
in the HDFS. The dashed line at $I=26$ mag represents the magnitude limit
for morphological classification.}
   \label{fig1}
   \end{center}
\end{figure}

\section{MORPHOLOGICAL CLASSIFICATION  AND NUMBER COUNTS}

\subsection{Method}

Morphological characteristics of faint galaxies provide critical information
for understanding the evolution of galaxies. However, it is very difficult
to classify galaxies in the HDF, and the results published so far are diverse 
(see Ferguson et al. 2000).  
We have classified the bright galaxies with $I<26$ mag in the HDFS
as well as in the HDFN.
For morphological classification we have used visual classification supplemented
with inspection of the surface brightness profiles of the objects. 
We have fit
the surface brightness profiles of the objects 
with de Vaucouleurs law, exponential law and King models. 
Fig. 2 displays the surface brightness profiles of typical
early-type and late-type galaxies.

\begin{figure}[ht] 
   \begin{center}
   \includegraphics[angle=0,width=8cm,clip]{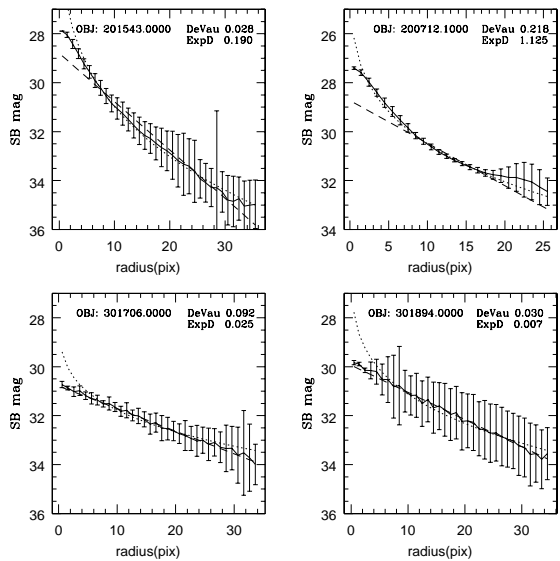}
   \caption{Surface brightness profiles of a sample of early-type (upper panels) and 
late-type (lower panels) galaxies.
The dotted line and the dashed line represent fits to the data 
with the de Vaucouleurs law and exponential law, respectively. 
One pixel in radius corresponds to 0.0398 arcsec.}
   \label{fig2}
   \end{center}
\end{figure}

\subsection{Results}

One of the well-known difficulties 
in classification of faint galaxies in the HST images
is due to splitting of a single galaxy into several bodies or merging of a few galaxies 
into one body by the softwares used. 
No known softwares including FOCAS and SExtractor can take care
of this problem perfectly. We have inspected each image visually to handle this
problem. It is very time and labor-consuming, but provides reasonably reliable
results.
 
Finally, we have classified about 340 galaxies in the HDFS and about 400 galaxies
in the HDFN into four classes: early-type, late-type (spirals), peculiar/irregular,
and merger. The classification schemes are basically similar to those used by
van den Bergh et al. (1996), but we used surface photometry of galaxies as well
for classification.
A comparison of our results for the HDFN with those of van den Bergh et al. shows
a good agreement between the two.

Fig. 3 illustrates the morphological population statistics for the classified
objects in the HDFS in comparison with those in the HDFN.
It is found that galaxy population statistics are similar between the HDFS and the
HDFN: the most dominant population in the HDFS is peculiar/irregular galaxies, and 
the late type galaxies are somewhat more than the early type galaxies. 
However, the number of stars in the HDFS is about twice 
that in the HDFN, which is consistent with the result given by 
Mendez \& Minniti (2000). It is as expected, because the HDFS is closer to the
center of our Galaxy than the HDFN.

\begin{figure}[ht] 
   \begin{center}
   \includegraphics[angle=0,width=8cm,clip]{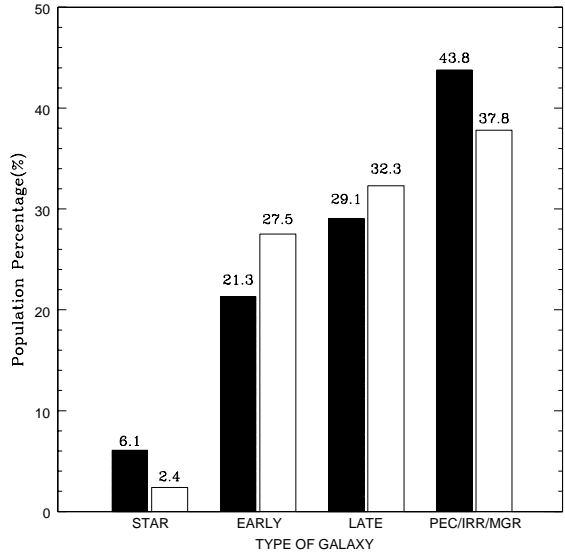}
   \caption{Morphological population statistics of the objects with $I<26$ mag
in the HDFS (black bars) and the HDFN (white bars). }
   \label{fig3}
   \end{center}
\end{figure}

\begin{figure}[hb] 
   \begin{center}
   \includegraphics[angle=0,width=8cm,clip]{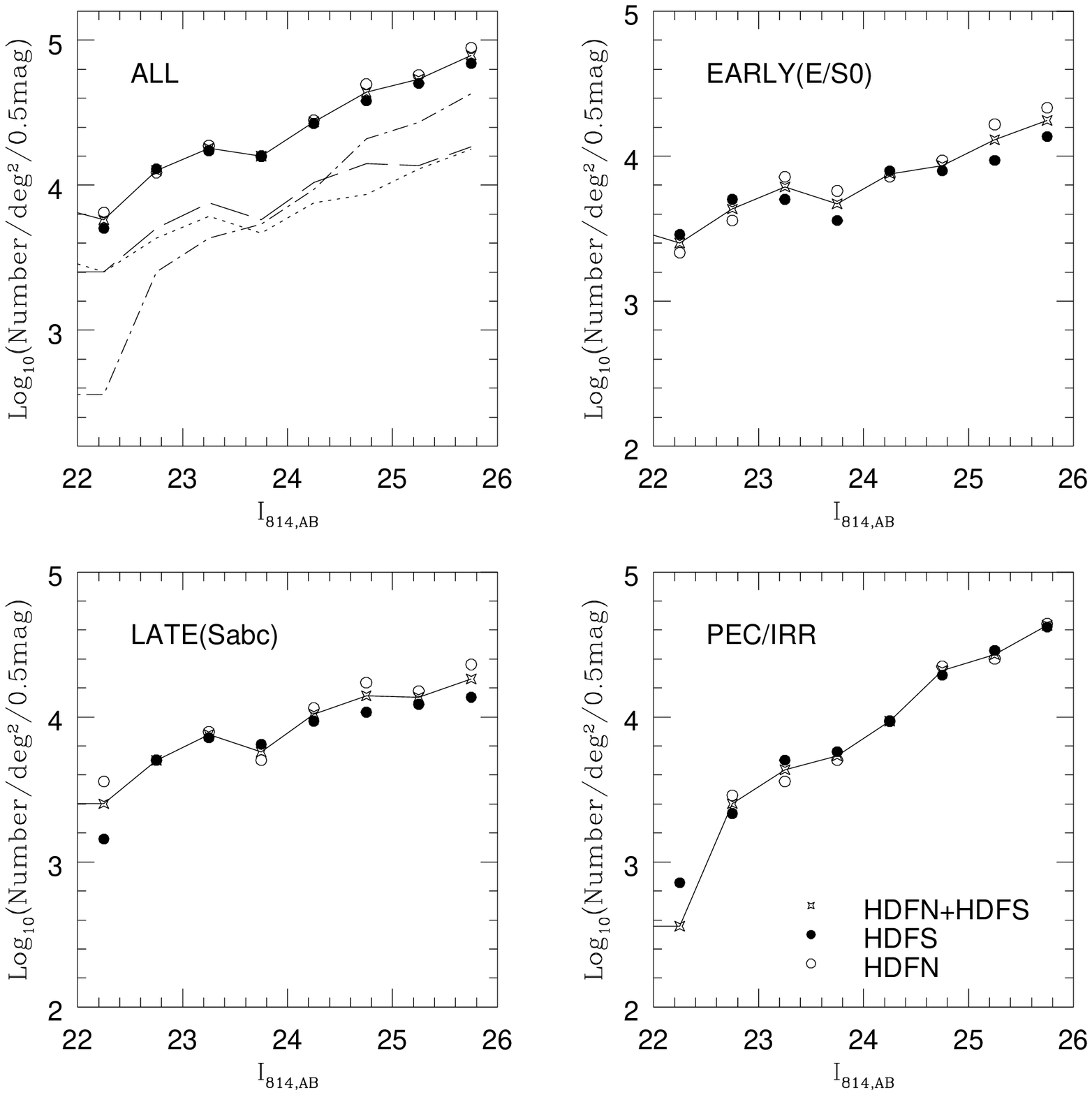}
   \caption{Morphological number counts at $I$-band of 
the galaxies with $I<26$ mag  in the HDFS (the filled circles)
and the HDFN (the open circles). 
The solid line represents the mean of the HDFS and the HDFN. 
The long-dashed line, short-dashed line and dot-dashed line in the left upper
panel represent, respectively, the late-type, early-type and peculiar/irregular galaxies.}
   \label{fig4}
   \end{center}
\end{figure}

We have then derived the $I$-band number counts of each type of galaxies 
with $I<26$ mag, and they are displayed in Fig. 4.
Fig.4 shows that the  number counts for the HDFS agree in general with those for
the HDFN, and that the number counts of the peculiar/irregular galaxies are steeper
than those of the early-type and late-type galaxies both of which are similar
to each other. The number of the peculiar/irregular galaxies is larger than
those for the other types at $I>24.5$ mag. Both of the early-types and late-types
in the both fields show a slight dip at $I\sim 23.8$ mag, which needs further investigation.

\section{PHOTOMETRIC REDSHIFTS}

We have determined photometrically the redshifts  of about 340 galaxies 
with $I<26$ mag in the HDFS and of about 400 galaxies in the HDFN 
using the empirical training set method described in Wang et al. (1998, 1999).
Wang et al. presented the calibration for the empirical training set
 based on 82 galaxies with measured spectroscopic redshifts $0.1<z<3.5$ 
in the HDFN.
Some galaxies detected in the master image are too faint to be detected in the
$U$ image. A value of $U=30$ mag was assumed as an upper limit 
for such $U$-band limited galaxies 
in determining the photometric redshift so that the redshifts of these galaxies
are more uncertain than those for the other galaxies.

Fig. 5 displays the redshift distribution of all the galaxies with $I<26$ mag
in the HDFS in comparison with that for the HDFN, and
Fig. 6 illustrates the redshift distribution of each type of the galaxies 
with $I<26$ mag.
A comparison of our results with those given for the galaxies with $I<26$ mag
in the HDFN given by Driver et al. (1998)
shows a reasonably good agreement between the two.

\begin{figure}[ht] 
   \begin{center}
   \includegraphics[angle=0,width=8cm,clip]{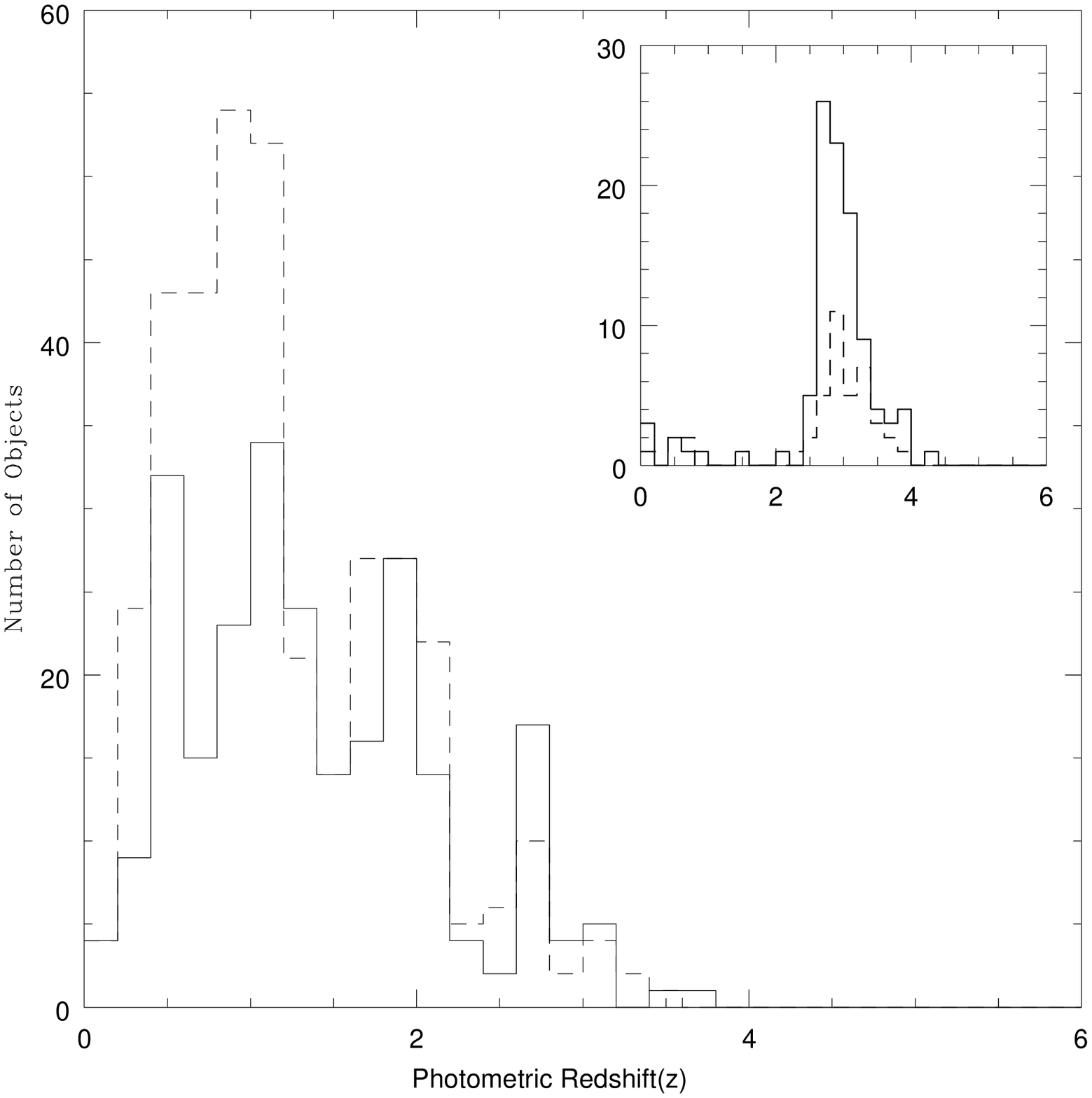}
   \caption{Photometric redshift distribution of the galaxies 
with $I<26$ mag in the HDFS (the solid line) and the HDFN (the dashed line).
The inlaid box represents the redshift distribution of the $U$-band limited
galaxies whose redshift estimates are more uncertain than for others. }
   \label{fig5}
   \end{center}
\end{figure}

\begin{figure}[htb] 
   \begin{center}
   \includegraphics[angle=0,width=14cm,clip]{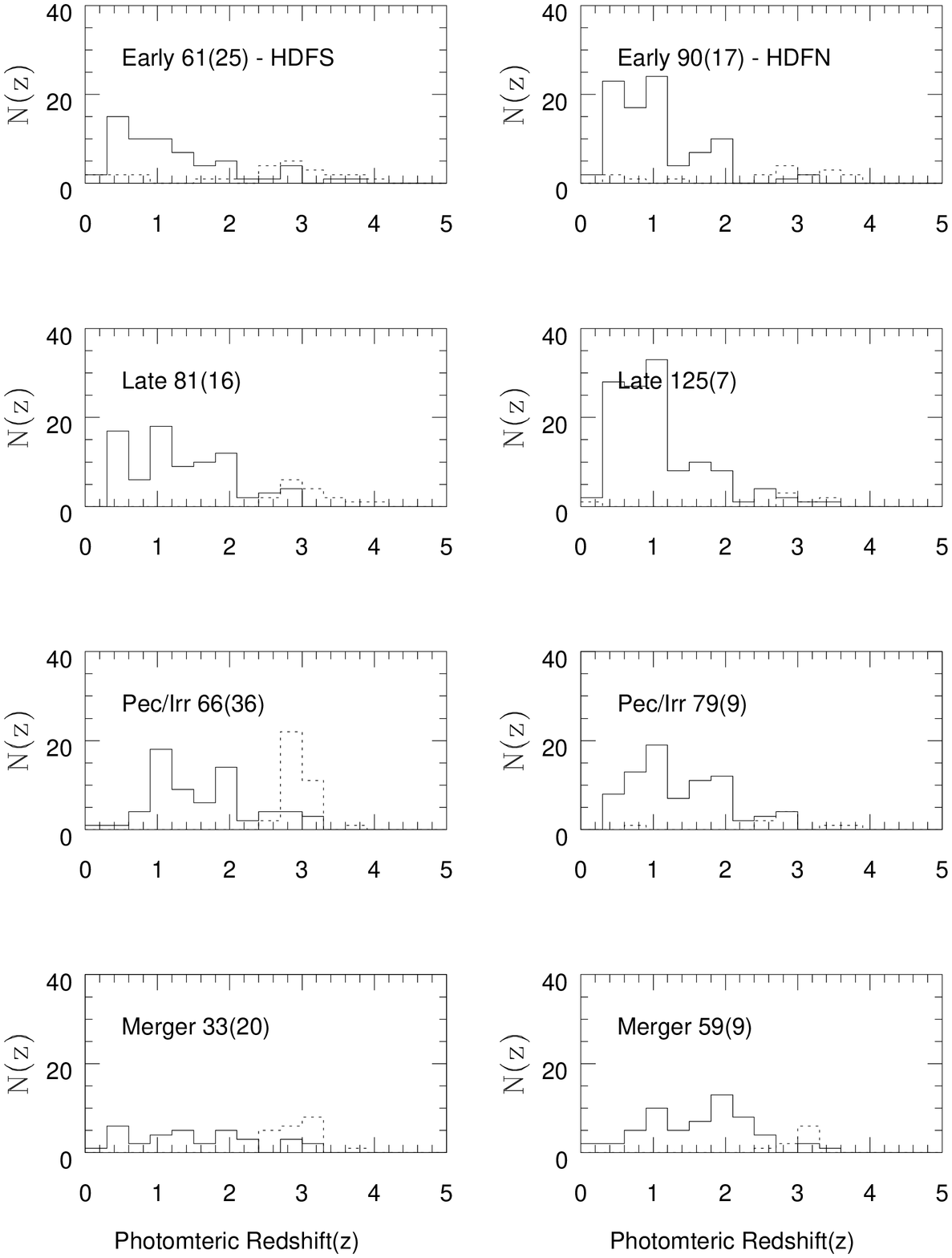}
   \caption{Morphological redshift distribution of the galaxies with $I<26$ mag 
in the HDFS (left) and the HDFN(right). The dotted lines represent the
$U$-band limited galaxies and the numbers inside the parentheses represent
the numbers of these galaxies.}
   \label{fig6}
   \end{center}
\end{figure}

Several distinguishable features are seen in Figs. 5 and 6:
a) The HDFS data show three strong peaks, 
at $z \sim$ 0.5, 1.0, 1.9, and one weaker peak at  $z \sim2.7$, 
while the HDFN data show one prominent peak at $z\sim 0.8$, 
one weak peak at $z \sim 1.8$, and a marginal peak at $z\sim 2.7$.   
The difference in the strongest peak between the HDFS and HDFN is due to
the stronger concentration of early-type and late-type galaxies within $z\sim1.2$
in the HDFN as compared to the HDFS, as shown in Fig. 6; 
b) The numbers of early-type and late-type galaxies in the HDFN decrease suddenly
at $z\sim1.2$ and to a less degree at $z\sim 2.1$, while those for the HDFS decrease more gradually with increasing redshift.
This, with the point in a), indicates that
the evolutionary history of the galaxies in the
HDFN and the HDFS might have been different; and
c) the merger and peculiar/irregular type galaxies show, on an average,  
larger redshifts than those of the late-type and early-type galaxies, which is seen
more clearly in the HDFN than in the HDFS. This shows that
 mergers were more frequent before the main population of 
the late-type and early-types.

Fig. 7 displays morphological color--redshift diagrams of the galaxies 
with $I<26$ mag in the HDFS. The colors vary significantly with increasing redshift.
For example, $(V-I)$ colors get redder for $0<z<1.0$, show a peak at $z\sim1.0$,
get bluer for $1.0<z<2$, and get redder again for $z>2.5$.
One interesting point seen in Fig. 7 is that the color range of early-type galaxies
at $z<1$ in the HDFS (and the HDFN) is very large from $(B-V)\sim0.0$ to $\sim 1.8$,
suggesting that the star formation histories in early-type galaxies have been
very diverse.  

\begin{figure}[ht] 
   \begin{center}
   \includegraphics[angle=0,width=9cm,clip]{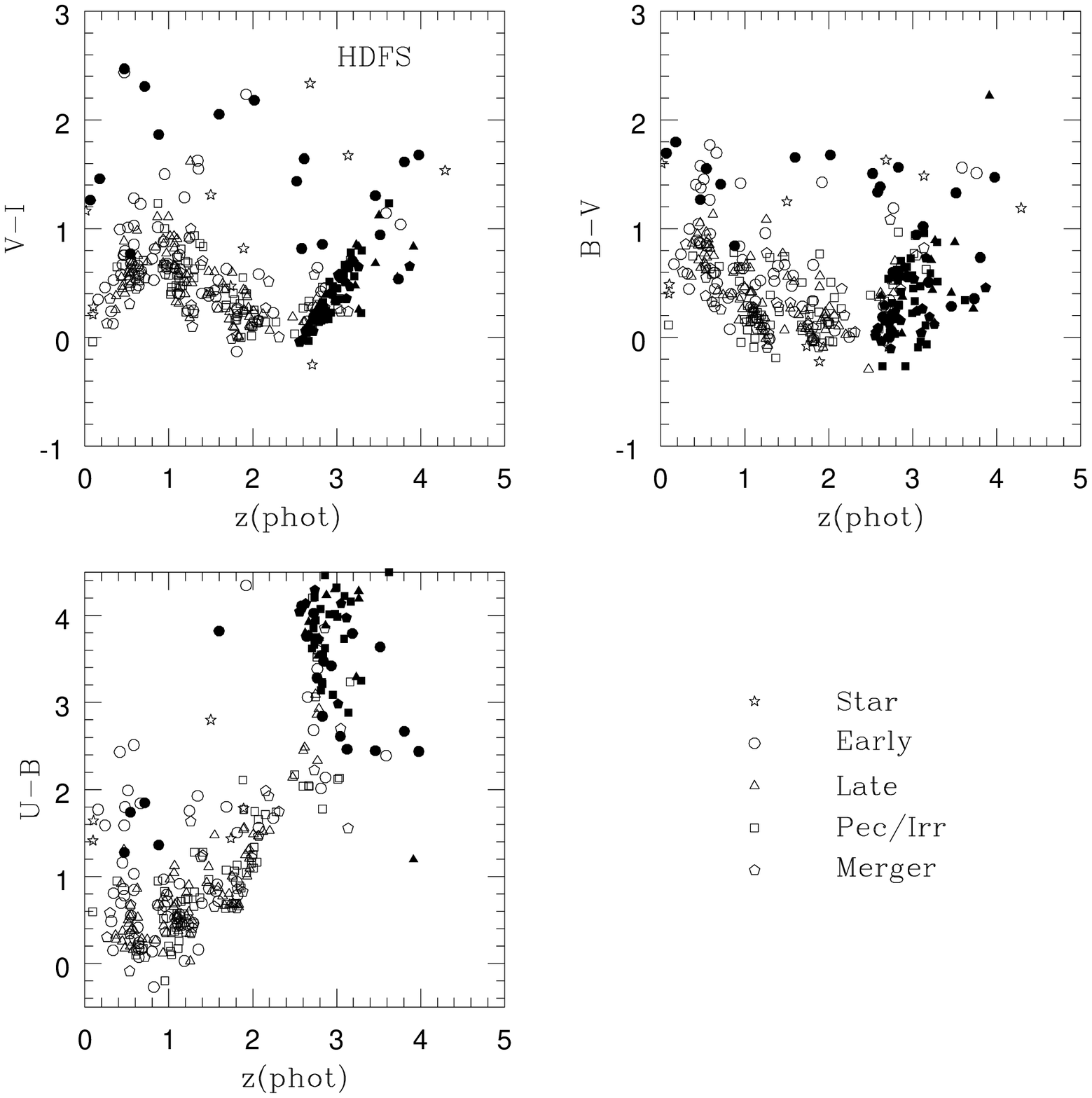}
   \caption{Morphological color--redshift diagrams of the galaxies with $I<26$ mag
in the
HDFS. 
Filled symbols represent the $U$-band limited galaxies.}
   \label{fig7}
   \end{center}
\end{figure}

Fig. 8 displays morphological magnitude--redshift diagrams of the galaxies 
with $I<26$ mag in the HDFS. Fig. 8 shows, in general, a distance effect on the redshift, i.e., 
the brightest galaxies get fainter with increasing redshift. However, 
there appears to be a turn-over in this trend around $z\sim 3$ (even excluding
the $U$-band limited galaxies). It will be interesting to confirm the redshifts
of the brightest galaxies at $z\sim 3$.
 
\begin{figure}[ht] 
   \begin{center}
   \includegraphics[angle=0,width=9cm,clip]{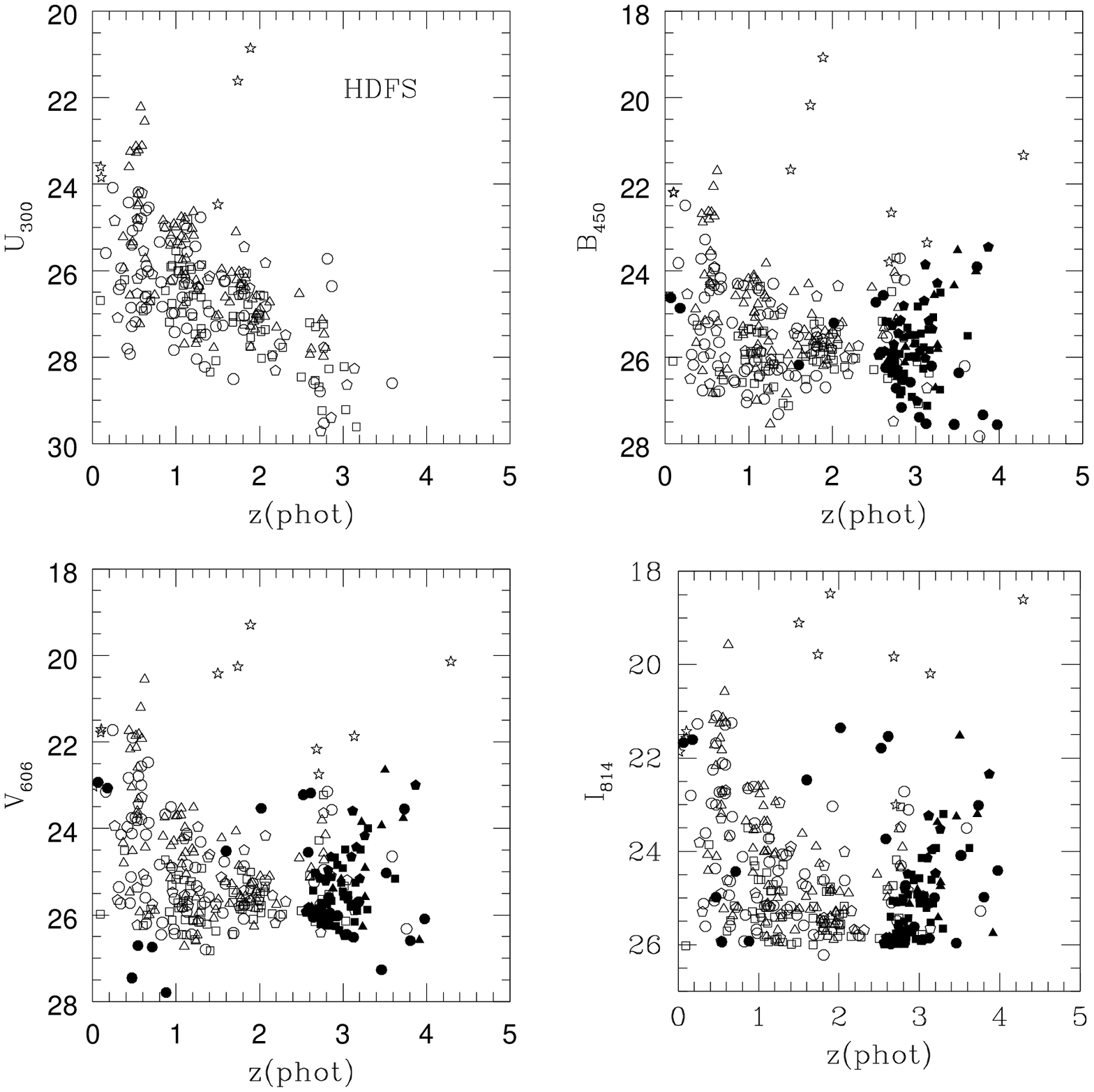}
   \caption{Morphological magnitude--redshift diagrams of the galaxies with $I<26$ mag
in the
HDFS. 
Filled symbols represent the $U$-band limited galaxies.
Open symbols are the same as in Fig. 7.}
   \label{fig8}
   \end{center}
\end{figure}

\section{SUMMARY}

We have investigated several photometric properties of the galaxies with $I<26$ mag
in the HDF in comparison with the HDFN. 
We have applied the same procedures for both the HDFS and HDFN for direct comparison,
so that the relative differences  we find between the two fields must be reliable. 
Next step will be to interpret the results by modelling
the formation and evolution of galaxies.
Since the invaluable data for the HDFN and HDFS were released, 
many papers have been published using these data. However,
there are still significant differences and/or conflicts as well as
agreements among the results (even among the data, excluding any models), 
showing clearly how difficult it is to mine the truth in the field
(see Ferguson et al. 2000 and references therein).
This indicates that other kinds of data in addition to the HDF data are needed
for achieving the final goal of understanding the formation and evolution 
of galaxeis. In this sense HDFS and HDFN are also very interesting targets 
for the coming infrared misssions including ASTRO-F and H2L2.





\section*{ACKNOWLEDGMENT}  
This research is in part supported by the BK21 program of the Korean Government.


\end{document}